%% file: main.tex
\documentclass[conference]{IEEEtran}
\IEEEoverridecommandlockouts

\usepackage{cite}
\usepackage{amsmath,amssymb,amsfonts}
\usepackage{algorithmic}
\usepackage{graphicx}
\usepackage{textcomp}
\usepackage{xcolor}
\usepackage[hyphens]{url}
\usepackage{comment}
\usepackage[caption=false,font=footnotesize]{subfig}

\usepackage{xspace}
\usepackage{hhline}
\usepackage{makecell}
\usepackage[utf8]{inputenc}
\usepackage[english]{babel}
\usepackage{xfrac}
\usepackage{booktabs}
\usepackage{enumitem}
\usepackage{color,soul}
\usepackage{multirow}
\usepackage{array}
\usepackage{placeins}
\usepackage{threeparttable}
\usepackage{siunitx}
\usepackage{stackengine}
\usepackage{listings}

\lstdefinestyle{aecommand}{
  basicstyle=\ttfamily\footnotesize,
  breaklines=true,
  breakatwhitespace=false,
  columns=fullflexible,
  keepspaces=true,
  showstringspaces=false,
  frame=tb,
  framerule=0.4pt,
  rulecolor=\color{black},
  xleftmargin=0pt,
  xrightmargin=0pt,
  aboveskip=1em,
  belowskip=1em
}

\newcommand{\ADD}[1]{\textcolor{black}{#1}}

\usepackage{tikz}
\newcommand{\circnumred}[1]{%
  \tikz[baseline=(char.base)]{
    \node[
      shape=circle,
      draw=red, fill=red,
      text=white,
      inner sep=.5pt
    ] (char) {#1};
  }%
}
\newcommand{\circnumorange}[1]{%
  \tikz[baseline=(char.base)]{
    \node[
      shape=circle,
      draw=orange, fill=orange,
      text=white,
      inner sep=.5pt
    ] (char) {#1};
  }%
}

\def\BibTeX{{\rm B\kern-.05em{\sc i\kern-.025em b}\kern-.08em
    T\kern-.1667em\lower.7ex\hbox{E}\kern-.125emX}}

\begin{document}
\bstctlcite{IEEEexample:BSTcontrol}
\pdfpagewidth=8.5in
\pdfpageheight=11in

\newcommand{\iscasubmissionnumber}{2023}
\newcommand{\mytitle}{RangeGuard}


\pagenumbering{arabic}

\title{RangeGuard: Efficient, Bounded \\Approximate Error Correction for Reliable DNNs}

\author{
\IEEEauthorblockN{
Hanum Ko\IEEEauthorrefmark{1},
Sangheum Yeon\IEEEauthorrefmark{1},
Jong Hwan Ko\IEEEauthorrefmark{2},
and Jungrae Kim\IEEEauthorrefmark{2}
}
\IEEEauthorblockA{
Sungkyunkwan University, Suwon, Republic of Korea\\
\IEEEauthorrefmark{1}Dept. of Semiconductor Convergence Engineering
\quad
\IEEEauthorrefmark{2}Dept. of Electrical and Computer Engineering\\
\{rhgksdma, shhj9787, jhko, dale40\}@skku.edu
}
}

\maketitle


\input{sections/0.abstract}




\input{sections/1.introduction}

\input{sections/2.background}

\input{sections/3.motivation}

\input{sections/4.prior}

\input{sections/5.main}

\input{sections/6.evaluation}

\input{sections/7.conclusion}



\bibliographystyle{IEEEtran}
\bibliography{references}


\end{document}

%% file: sections/0.abstract.tex
\begin{abstract}

As DRAM scales in density and adopts 3D integration, raw fault rates increase and multi-bit errors are no longer rare. Such errors can severely impact \emph{Deep Neural Networks (DNNs)}: although DNNs tolerate small numerical perturbations, random bit flips can create extreme outliers that propagate and sharply degrade accuracy. \emph{Large Language Models (LLMs)} are particularly vulnerable because attention, residual, and normalization layers can amplify and preserve a single corrupted activation across many layers, destabilizing inference.

This paper introduces \emph{\mytitle{}}, a metadata-centric error-correcting framework that provides strong reliability and high efficiency based on \emph{bounded approximate correction}. Instead of protecting raw bits, \mytitle{} encodes compact \emph{Range Identifiers (RIDs)} that capture the numerical range of each value.
These compact metadata enable efficient use of limited redundancy and concentrate protection on range changes—which indicate harmful semantic deviations—while ignoring benign intra-range variations. Upon detecting a range change, \mytitle{} restores the correct range and substitutes a representative value, ensuring that error magnitudes are bounded within the range.
Based on RIDs, \mytitle{} can tolerate 64+ bits of error using only 16 bits of parity available in GPU memories without a noticeable accuracy loss.
By introducing semantic range protection, \mytitle{} enables reliable DNN execution even under frequent memory errors and tight redundancy budgets.


\end{abstract}

%% file: sections/1.introduction.tex
\section{Introduction}
\label{sec:intro}

DRAM reliability is degrading with density/voltage scaling and 3D integration. Conventional 2D devices (e.g., DDR4/5)~\cite{standard2021ddr4,standard2022ddr5} face shrinking noise margins and rising marginal cell defects. \ADD{High Bandwidth Memory (HBM)} further introduces complex processing (e.g., wafer thinning, \ADD{Through Silicon Via (TSV)} formation, thermal cycling) and harsher thermal/mechanical stress, which elevate fault susceptibility. Fleet\mbox{-}scale reports find HBM2 raw error rates substantially higher than DDR4 ($\sim500\times$), and production GPU clusters observe uncorrectable multi\mbox{-}bit HBM3 errors on the order of hours~\cite{li2022correctable,wu2024removing}. As systems provision more memories for bandwidth and capacity, dense multi\mbox{-}bit faults are no longer rare\mbox{-}event tails.

These faults can be devastating for \emph{Deep Neural Networks (DNNs)}. While DNNs tolerate small numerical perturbations—such as quantization noise or rounding error that remains a tiny fraction of the original value—random \emph{bit flips} can cause catastrophic deviations. For example, flipping the most significant exponent bit of a \emph{bfloat16 (BF16)} value turns \(0.5\) into \(2^{127}\!\approx\!1.7\times10^{38}\), injecting an astronomically large outlier that can propagate through the network and sharply degrade accuracy.

\emph{Large Language Models (LLMs)} are particularly susceptible. 
Unlike \emph{Convolutional Neural Networks (CNNs)}, where spatial filters may attenuate local anomalies, a transformer in an LLM can \emph{amplify} a single corrupted token feature through multi-head attention, \emph{preserve} it across layers via residual connections, and \emph{stabilize} its presence through normalization. This combination lets one faulty activation persist, interact with many tokens, and cascade through subsequent blocks, magnifying its impact on logits and sequence generation, and ultimately destabilizing the final output (Section~\ref{sec:motivation:dnn_accuracy}).

However, providing robust protection under realistic system constraints is challenging. 
DNN accelerators (e.g., GPUs) operate on small memory-access granularities (e.g., 32 bytes) to avoid overfetch, leaving little opportunity to amortize redundancy. 
Contemporary GPU memories provision only 2 bytes of parity per 32-byte block (6.25\%), yet practical failure modes can corrupt up to \emph{4 bytes} within the same block~\cite{ryu202316}. 
Under this tight budget, conventional \emph{bit-centric} protection schemes struggle to handle dense multi-bit corruptions without incurring unacceptable storage or bandwidth overheads. 
These constraints motivate rethinking how to spend scarce parity so that it protects what actually matters for model correctness.

This paper introduces \emph{\mytitle{}}, a metadata-centric error-correcting framework for DNNs and approximate computing. 
Instead of protecting raw bits, \mytitle{} protects per-value \emph{metadata}: a compact \emph{Range Identifier (RID)} that encodes the value’s numeric interval. 
This design enables far more efficient use of limited redundancy: for example, restoring a corrupted 32-bit value from a dense 32-bit fault can require at least 64 bits of redundancy, whereas restoring a 4-bit RID of that value needs only 8 bits. 
Moreover, \mytitle{} allocates its correction capability to errors that \emph{change ranges}. 
If bit flips perturb only lower bits and keep the value within the same range, the RID is unchanged and consumes no correction budget. 
Only severe errors that flip critical bits and cause an \emph{inter-range} deviation trigger correction, ensuring that scarce redundancy is spent on semantically harmful faults rather than benign intra-range perturbations.

\begin{figure}[t]
    \centering
    \includegraphics[width=0.712\columnwidth]{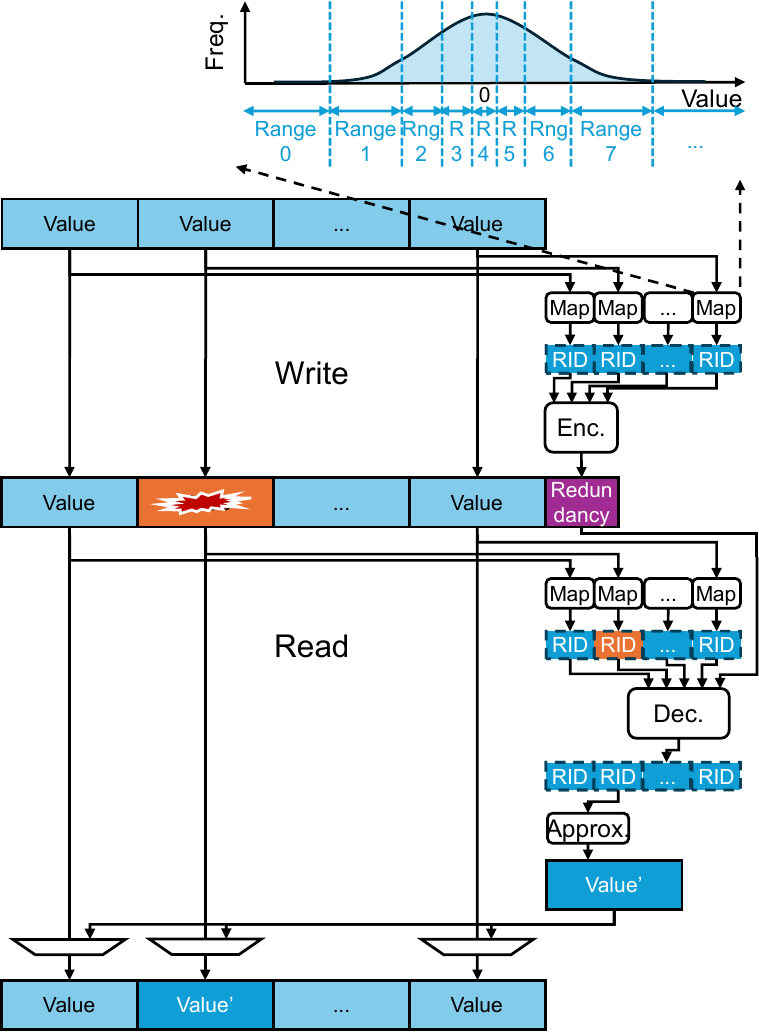}
    \vspace{-5pt}
    \caption{An overview of \mytitle{}. ECC operates on RIDs, rather than raw data bits, and repairs corrupted data by restoring it to a representative value of the range.}
    \label{fig:intro}
    \vspace{-15pt}
\end{figure}

Figure~\ref{fig:intro} presents an overview of \mytitle{}. 
On a write, each value is mapped to an RID via a \ADD{\emph{RangeMap}}, and an \emph{Error Correcting Code (ECC)} encodes the RID symbols into a short redundancy field. 
The explicit RIDs are then discarded; only the original data and the encoded RID redundancy are stored in memory. 
On a read, \mytitle{} regenerates candidate RIDs from the fetched data and applies ECC decoding over these regenerated RIDs and the stored redundancy. 
Inter-range errors are detected and corrected by the decoder, whereas intra-range errors remain undetected and do not consume correction capacity. 
After restoring the RID, \mytitle{} replaces the corrupted value with a representative value for that range, so the resulting error magnitude is bounded by the range width and has limited impact on DNN accuracy.

\ADD{Because each RangeMap is tailored to a specific data format and value distribution, a single map offers limited flexibility in modern DNN systems with multiple formats and diverse value distributions. To address this issue, \mytitle{} utilizes multiple maps and address-based ECC selection~\cite{yoon2010virtualized}. The memory controller maintains multiple maps for different numeric formats or value distributions while also preserving a conventional exact-correction scheme. The operating system supports the appropriate protection mode for each data region by encoding ECC-selection metadata (\emph{Map Tag}) in otherwise unused upper physical-address bits (e.g., bits [57:54]). Since these bits lie beyond the implemented physical-memory capacity, they do not affect the actual memory location and instead serve only as control metadata. This mechanism enables flexible per-region protection without disrupting normal address translation or existing ECC operation.}

Our evaluation shows that \mytitle{} delivers strong robustness to dense, multi-bit corruptions within the standard redundancy budget. 
Using the 2-byte redundancy per 32-byte block available in contemporary GPU memories, one \mytitle{} instantiation tolerates two inter-range errors affecting up to 64 data bits, while bypassing an unbounded number of intra-range errors. 
This strong protection preserves LLM accuracy even at extremely high \emph{Bit Error Rates (BERs)}: for example, an unprotected Llama-3.2-1B model suffers significant accuracy loss from just a single bit flip at critical positions, whereas \mytitle{} maintains robust execution up to BER $=10^{-6}$ and incurs only small degradation at BER $=10^{-5}$. 
At the same time, \mytitle{} operates within existing GPU memory interfaces and incurs negligible performance overhead and no additional storage overhead.

The major contributions of this paper are as follows:
\begin{itemize}
    \item We propose \mytitle{}, a metadata-centric protection framework that uses per-value Range Identifiers (RIDs) to provide strong, efficient reliability under tight redundancy budgets.
    \item We analyze how DRAM bit flips map to value errors and quantify how these errors degrade end-to-end accuracy for both CNNs and LLMs, revealing LLMs are highly vulnerable to errors possibly due to their transformer layers.
    \item We develop an optimization framework for constructing the value-to-RID range mapping and present practical heuristics that minimize its overhead while preserving protection quality.
    \item We design a memory system that supports diverse data formats and value distributions with \mytitle{}.
\end{itemize}

The rest of this paper is organized as follows.
Section~\ref{sec:background} provides background on numeric formats and ECC.
Section~\ref{sec:motivation} characterizes how bit flips translate into value errors and end-to-end accuracy loss.
Section~\ref{sec:prior} discusses prior work on DNN fault tolerance and ECC.
Section~\ref{sec:main} presents the design of \mytitle{}, including RangeMap construction and architectural support.
Section~\ref{sec:eval} evaluates \mytitle{}'s reliability and overheads.
Section~\ref{sec:conclusion} concludes the paper.

%% file: sections/2.background.tex
\section{Background}
\label{sec:background}

This section lays the groundwork for \mytitle{} by reviewing DNN data types and ECC fundamentals.

\subsection{DNN Data Types}
\label{sec:data_type}

DNNs trade accuracy, throughput, memory footprint, and energy through their numeric formats. Two families dominate: \emph{floating point} (for wide dynamic range and optimization stability) and \emph{integer} (for efficient deployment via quantization).

\subsubsection{Floating-point \ADD{(FP)}}
Floating-point formats capture fine-grained value changes while offering large dynamic range. Although FP32 \ADD{(32-bit)} was the historical default, modern training is largely mixed precision: FP16 \ADD{(16-bit)} or BF16 \ADD{(\emph{bfloat16}, 16-bit)} for activations/gradients, with FP32 accumulators or “master” weights.
\ADD{FP16 typically employs dynamic loss scaling to avoid underflow, whereas BF16 retains FP32's exponent range to reduce the risk of underflow and improve training stability. Emerging 8-bit floating-point formats (FP8) typically use 4-bit or 5-bit exponents (E4M3 and E5M2, respectively), enabling lower storage, memory traffic, and computation cost.}

\subsubsection{Integers \ADD{(INT)}}
After training, many systems \emph{quantize} tensors to low-bit \emph{integers} for efficient inference. Quantization maps real values to integers via a scale (and optionally a zero point). A widely used baseline employs INT8 weights and activations with INT32 accumulation; pushing below 8 bits (e.g., INT4/INT3) generally requires quantization-aware training and related techniques to preserve accuracy.

Quantization leverages DNNs' tolerance to small perturbations: under uniform quantization, the absolute error is bounded by \ADD{the quantization step size and is typically} only a small fraction of the original value, yielding negligible accuracy loss. In contrast, a \emph{bit flip} can cause far larger deviation than the original value. For floating point, flipping exponent bit \(p\) \emph{scales} the value by \(2^{\pm 2^{p}}\), exhibiting an effectively \emph{double-exponential} sensitivity to bit position (Section~\ref{sec:motivation:value_errors}). Such deviations beyond the original value can be intolerable for DNNs (Section~\ref{sec:motivation:dnn_accuracy}), motivating strong protection against memory errors.

\subsection{Error Correcting Codes}
\label{sec:ecc}

Error Correcting Codes (ECC) can detect and correct errors by adding redundancy to data, forming a codeword~\cite{yoon2010virtualized,udipi2012lotecc,2015bambooecc,kim2016allinclusive,2023unity}. A common baseline is Single-Error Correction, Double-Error Detection (SEC–DED), which corrects any single-bit error and detects (but cannot correct) any double-bit error in a codeword.

DRAM errors often appear as \emph{bursts} aligned with the device’s internal organization (e.g., I/O pins, sub-wordline). In such settings, non-binary, \emph{symbol}-based ECC can be more efficient than bit-oriented codes: by grouping a spatially clustered error into one symbol, the decoder repairs the entire cluster as a single unit, reducing the redundancy needed for a given level of protection.

Reed–Solomon (RS)~\cite{reed1960polynomial} codes are a canonical family of symbol-based codes.
An RS$(n,k)$ code protects $k$ data symbols with $(n{-}k)$ parity symbols to form an $n$-symbol codeword. The decoder can correct up to $\left\lfloor (n-k)/2 \right\rfloor$ arbitrary symbol errors, provided the code length satisfies $n \le 2^m - 1$\ADD{, where $m$ is the symbol size}. In practice, memory word sizes rarely match the full code length, so systems use \emph{shortened} RS codes: some leading symbol positions are conceptually fixed (not stored/transmitted), and encoding/decoding proceed as usual over the remaining positions. Shortening preserves the same error-correction capability for the active symbols; corrections in the omitted positions are reported as error detections instead of silent miscorrections.

\mytitle{} uses two RS configurations:
(1) a 4-bit RS$(12,8)$, providing \emph{Double-Symbol Correction (DSC)} for 4-bit RID symbols, and
(2) an 8-bit RS$(10,8)$, providing \emph{Single-Symbol Correction (SSC)} for 8-bit RID symbols.
Both consume exactly 16 parity bits per 256-bit block, matching the HBM budget.

%% file: sections/3.motivation.tex
\section{Motivation}
\label{sec:motivation}

This section explains how low–level bit flips can propagate to large value deviations in DNN tensors and ultimately degrade model accuracy.

\subsection{HBM Errors}
\label{sec:motivation:errors}

HBM underpins modern AI infrastructure by delivering extreme bandwidth and high energy efficiency via TSVs and silicon interposers. The same 2.5D integration, however, introduces—and amplifies—reliability hazards. Dense stacking, TSV/microbump interconnects, and elevated operating temperatures increase the likelihood of manufacturing defects, thermo-mechanical stress, and heat-induced marginalities compared to planar DDR devices \cite{jun2017hbm,ha2023reliability,li2017reliability,chun202016,jun2016high,lee202413,
farmahini2018challenges,lee2014tsv,kim2023thermal}.

Fleet-scale measurements demonstrate a pronounced reliability gap between stacked HBMs and traditional DRAMs~\cite{li2022correctable,wu2024removing,beigi2023systematic, du2021predicting, llama2024hbm3}.
A useful comparison is HBM2 versus DDR4, because both expose \emph{raw} device errors (i.e., no on-die ECC). 
In ByteDance’s server fleet (100{,}000 servers with DDR4 memories), the average incidence over eight months is \emph{0.07 errors per device per month} \cite{li2022correctable}. In contrast, Huawei’s study of 15{,}000 domain-specific accelerators over two years reports a mean of \emph{35 errors per device per month} for HBM2 core dies \cite{wu2024removing}. While platform and workload differences preclude a perfectly controlled comparison \ADD{(HBM with higher bandwidth can exhibit more frequent errors from a permanent fault)}, the resulting \emph{$\sim$500$\times$} gap highlights the severity of HBM reliability challenges.

Later generations (e.g., HBM3 and DDR5) integrate \emph{on-die ECC (O-ECC)} that corrects many errors before they exit the bank group, reducing externally visible errors. Public descriptions indicate that HBM3 employs stronger device-internal coding than DDR5 (e.g., 16-bit symbol correction with 12.5\% internal redundancy versus single-bit correction with 6.25\% in DDR5), targeting multi-bit patterns that arise from peripheral faults~\cite{gurumurthi2021hbm3,park2022192,ryu202316}.

However, the scope of O-ECC is inherently limited: faults originating \emph{outside} the bank group—global data/command lines, I/O circuitry, and TSVs/microbumps—can still surface at the system boundary. Moreover, certain \emph{intra}-bank peripheral failures can exceed the 16-bit correction capability.
For example, a HBM3 sub-wordline (SWL) driver typically fans out to 32 data bits per access and operates at elevated voltages to suppress leakage~\cite{ryu202316}; this stress raises wear-out risk, and a single defective driver can corrupt all serviced bits in a burst. Such 32-bit errors overwhelm O-ECC’s correction capability and propagate to the host.

As a result, data centers continue to observe non-trivial \emph{exposed} HBM errors. During a 54-day Llama-3 (405B) pre-training run on 16{,}384 H100 GPUs, Meta reported \emph{72} unexpected job interruptions attributed to uncorrectable HBM3 errors~\cite{llama2024hbm3}.
The GPUs utilize processor-side ECC (\emph{System ECC, or S-ECC}) with Single Error Correction (SEC) capability, and this number indicates that the cluster experienced exposed multi-bit errors every 18 hours. 

Although differences in platforms and measured metrics (e.g., job interruptions versus error counts) prevent a perfectly controlled comparison, this reported failure rate represents a substantial improvement over HBM2 (from 5,600 FIT/device to 90 FIT/device). However, the overall uncorrectable error rate remains significantly higher than that of DDR4 (near zero FIT), underscoring the considerable reliability challenges in recent HBMs.

\subsection{Memory Errors to Value Errors}
\label{sec:motivation:value_errors}

Not all memory errors translate into the same numerical impact: the magnitude of a \emph{value error} depends on (i) the data type, (ii) flipped bit location, and (iii) the original value. We illustrate this sensitivity with single-bit flips on BF16, while \mytitle{} also considers multi-bit faults and other data types.

\begin{figure}[t]
    \centering
    \includegraphics[width=.8\columnwidth]{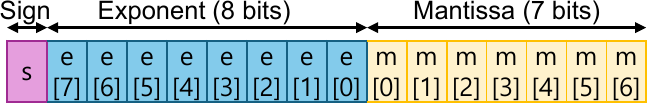}
    \caption{The bfloat16 (BF16) data format.}
    \label{fig:bf16}
    \vspace{-15pt}
\end{figure}

A BF16 number has one sign bit $\mathbf{s}$, eight exponent bits $\mathbf{e}$ (bias $=127$), and seven mantissa (fraction) bits $\mathbf{m}$ (Figure~\ref{fig:bf16}). For a normalized value,
\begin{equation}
x \;=\; (-1)^s \times 2^{e-\text{bias}} \times (1.m),
\end{equation}
where $(1.m)$ denotes the implicit leading-$1$ mantissa.

\subsubsection{Sign-bit flip}
Toggling the sign ($s \!\to\! 1-s$) yields $x'=-x$ and an absolute error $|x'-x|=2|x|$. This error can be large but is \emph{bounded} and strictly proportional to the original magnitude.

\subsubsection{Mantissa-bit flip}
Flipping the $k$-th mantissa bit (weight $2^{-(k+1)}$, with $k{=}0$ the most significant) produces

\[
x'=(-1)^s2^{e-\text{bias}}\bigl((1.m)\pm2^{-(k+1)}\bigr)
=x\!\left(1\pm\frac{2^{-(k+1)}}{1.m}\right)
\]

\begin{equation}
|x'-x| \;=\; \frac{2^{-(k+1)}}{1.m} \times |x| \leq \frac{1}{2} |x|.
\end{equation}

Thus a \emph{single} mantissa-bit flip cannot exceed half the original magnitude (tight when $m{=}0$, $k{=}0$), and its effect decays exponentially with $k$. Because these contributions form a geometric series, the cumulative perturbation from mantissa errors remains bounded by the original value (i.e., $|x'-x|<|x|$) even if \emph{all} mantissa bits flip.

\subsubsection{Exponent-bit flip} 
\begin{figure}[t]
    \centering
    \subfloat[Error ratio (=$x'/x$) under a single exponent-bit flip.]{
        \includegraphics[width=.8\columnwidth]{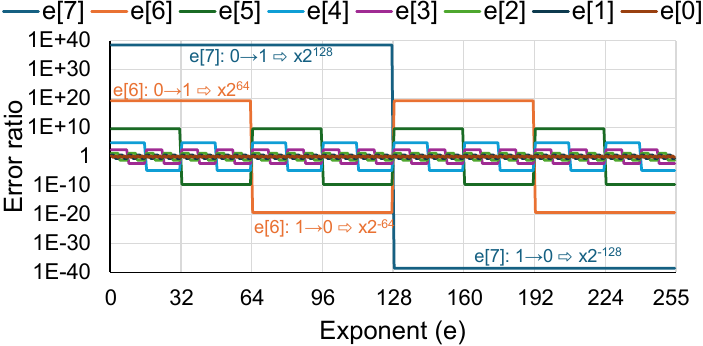}
        \label{fig:data_type_1}
        \vspace{-2pt}
    }\\
    \subfloat[Error magnitude (=$|x'-x|$) for the same scenarios.]{
        \includegraphics[width=.8\columnwidth]{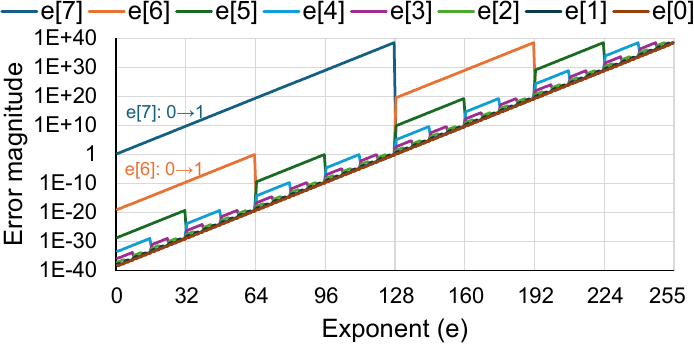}
        \label{fig:data_type_2}
    }\\
    \caption{Impact of exponent-bit flips in BF16. We vary the original exponent value $\mathbf{e}$ and its flip bit position $\mathbf{e[i]}$.}
    \label{fig:data_type}
    \vspace{-10pt}
\end{figure}

Exponent flips can be catastrophic. Flipping the $p$-th exponent bit (weight $2^{p}$; with $p=0$ the least significant) changes the stored exponent by $\Delta e=\pm 2^{p}$, giving
\begin{equation}
x' = (-1)^s \times 2^{\,e+\Delta e-\text{bias}} \times (1.m) = 2^{\Delta e} \times x.
\end{equation}
This has three implications. First, the \emph{Scale Factor ($\mathrm{SF}=2^{\Delta e}=2^{\pm 2^p}$)} grows \emph{double-exponentially} with bit index, so high-order flips cause orders-of-magnitude shifts. Second, error \emph{polarity} matters: 
$0{\to}1$ amplifies ($\mathrm{SF} = 2^{+2^{p}} \ge 2$) and can yield effectively \emph{unbounded} error (e.g., $p{=}7$ gives $\mathrm{SF}=2^{128}\!\approx\!3\times 10^{38}$),
whereas $1{\to}0$ attenuates ($\mathrm{SF} = 2^{-2^{p}} \le \frac{1}{2}$), pulling the value towards zero and producing a \emph{bounded} error not exceeding $|x|$.

To visualize $\mathrm{SF}$s, we inject 
single-bit flips across different bit positions of specific exponent values $e$ and plot the error ratio (=$x'/x$) (Figure~\ref{fig:data_type_1}).
The result shows a dramatic range, spanning from \(2^{-128}\) to \(2^{128}\)—with alternating bands based on polarity (\(0 \to 1\) vs. \(1 \to 0\)).
For example, when $e{=}64$, flipping $e[7]$ ($64\!\to\!192$) multiplies the value by $2^{128}\!\approx\!3\times10^{38}$; conversely, flipping the same bit when $e{=}192$ ($\to 64$) yields $2^{-128}\!\approx\!3\times10^{-39}$, effectively driving the value toward zero. Finally, flips that produce $e{=}0$ or $e{=}255$ enter BF16’s special regimes (subnormals/zero or $\pm\infty$/NaN), causing discontinuities or NaNs (not shown).

\begin{table}[t!]
\centering
\caption{Summary of error ratios (=x'/x) induced by bit flips.}
\scriptsize
\setlength{\tabcolsep}{3pt}
\renewcommand{\arraystretch}{1.1}
\begin{tabular}{c|c|c|c|c|c|c|c|c|c}
\Xhline{2\arrayrulewidth}
\multirow{2}{*}{\textbf{Bit Flip}} & \multicolumn{9}{c}{\textbf{BF16 Bit Position}} \\

\cline{2-10}

& \textbf{$s$} & \textbf{$e[7]$} & $e[6]$ & $e[5]$ & ... & $e[0]$ & $m[0]$ & $m[1]$ & ... \\

\Xhline{2\arrayrulewidth}

$0\to1$ &
\multirow{2}{*}{$\times(-1)$} &
$\times2^{2^{7}}$ &
$\times2^{2^{6}}$ &
$\times2^{2^{5}}$ &
... &
$\times2^{2^{0}}$ &
\multirow{2}{*}{$\times2^{-1}$} &
\multirow{2}{*}{$\times2^{-2}$} &
\multirow{2}{*}{...} \\

\cline{1-1}\cline{3-7}

$1\to0$ &
&
$\times2^{-2^{7}}$ &
$\times2^{-2^{6}}$ &
$\times2^{-2^{5}}$ &
... &
$\times2^{-2^{0}}$ &
& & \\

\Xhline{2\arrayrulewidth}
\end{tabular}

\label{tab:bf16_summary}
\vspace{-10pt}
\end{table}

Table~\ref{tab:bf16_summary} summarizes BF16’s bit-level vulnerability. Exponent bits are highly fragile: their error ratios grow double-exponentially with bit index. Moreover, the effect is strongly polarity-dependent: $0{\to}1$ flips cause catastrophic amplification, whereas $1{\to}0$ flips induce large attenuation.
The sign always introduces a ratio of $-1$, but its impact is overshadowed by the extreme amplifications in exponent bits. In contrast, mantissa bits only introduce fractional changes, and their ratios decay exponentially with bit position, making them comparatively benign.

The final implication is that the \emph{error magnitude} depends on both the scale factor and the original value ($|x'-x| = |\mathrm{SF}-1|\cdot|x|$).
When $|x|$ is small, even large multiplicative factors can yield small absolute deviations; conversely, for large $|x|$, even small factors can produce large errors. Figure~\ref{fig:data_type_2} visualizes this behavior for exponent-bit flips with $m{=}0$: the error scales with the original value and exhibits clear polarity asymmetry—$0{\to}1$ (amplification) dominates $1{\to}0$ (attenuation).

\subsection{Memory Errors to DNN Accuracy}
\label{sec:motivation:dnn_accuracy}

\begin{figure*}[t]
    \centering
    \subfloat[ResNet-50 (CNN) / ImageNet-1K top-1 accuracy]{
        \includegraphics[width=.31\textwidth]{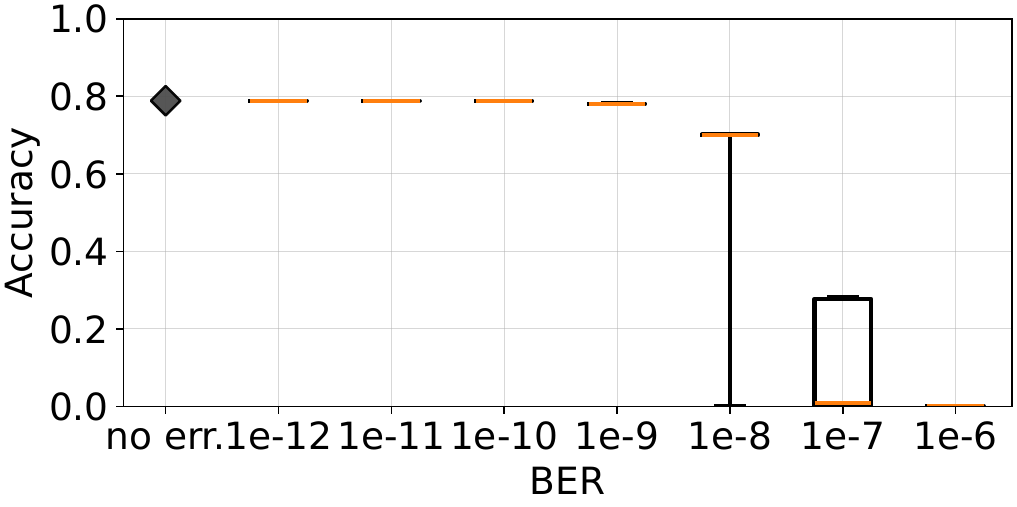}
        \label{fig:dnn_comparison_cnn}
    }
    \hspace{3pt}
    \subfloat[Llama-3.1-8B (LLM) / ARC-Easy score]{
        \includegraphics[width=.31\textwidth]{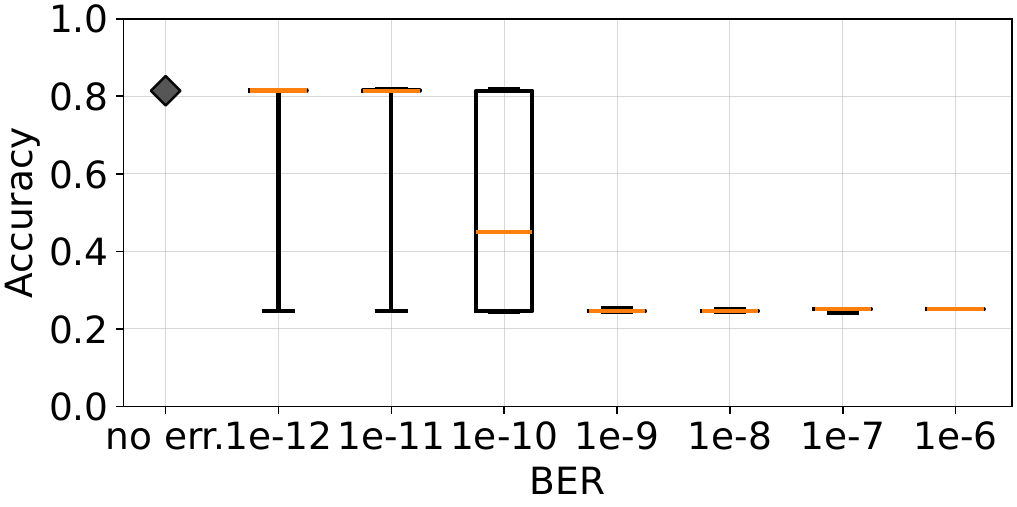}
        \label{fig:dnn_comparison_llm_8B}
    }
    \hspace{3pt}
    \subfloat[Llama-3.2-1B (LLM) / ARC-Easy score]{
        \includegraphics[width=.31\textwidth]{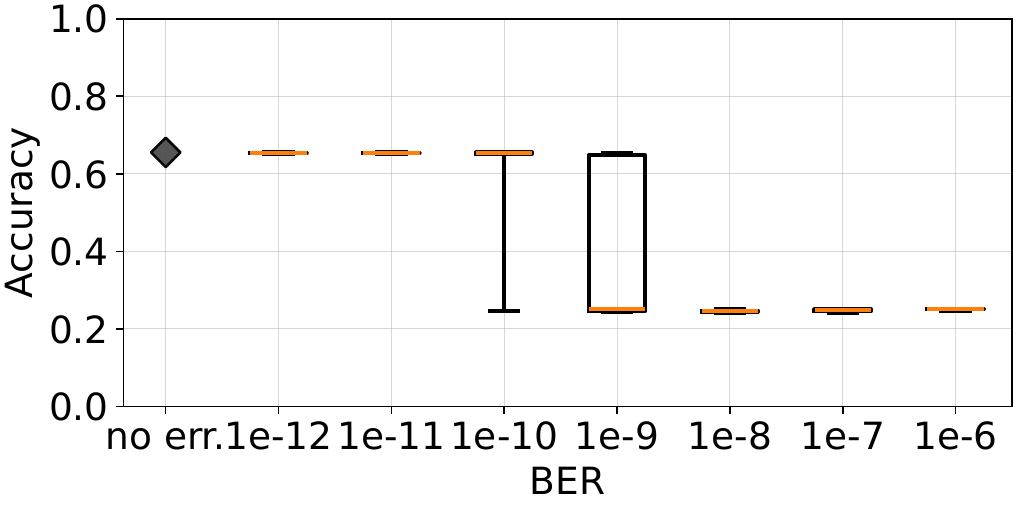}
        \label{fig:dnn_comparison_llm_1B}
    }
    \vspace{-10pt}
    
    \subfloat[Llama-3.2-1B (LLM) / ARC-Easy score under bit flips at specific bit positions.]{
        \includegraphics[width=0.75\textwidth]{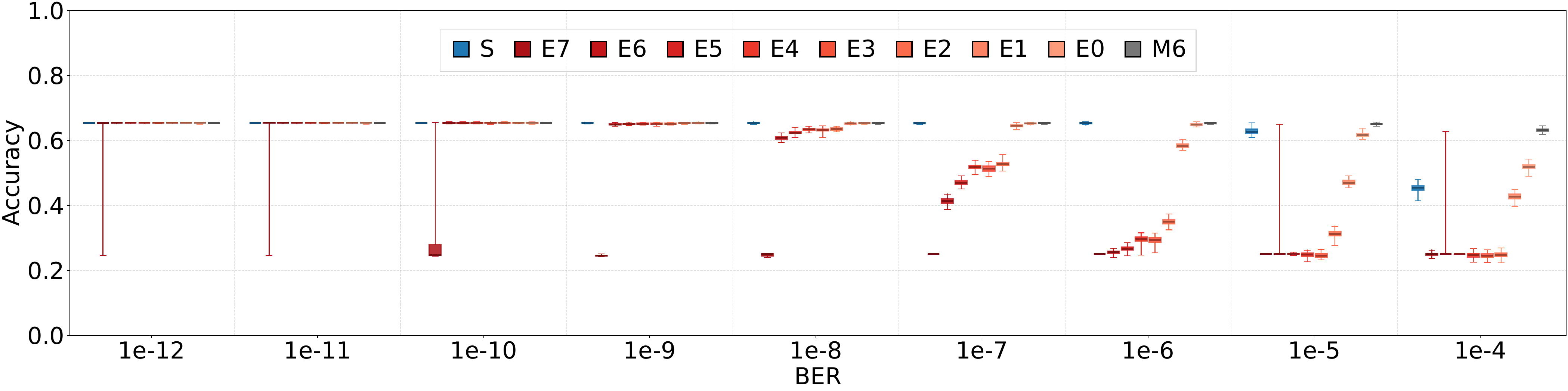}
        \label{fig:dnn_bitposition}
        \vspace{-10pt}
    }
    \caption{DNN accuracy under random bit flips across varying BERs.}
    \label{fig:dnn_comparison}
    \vspace{-15pt}
\end{figure*}

We next examine how these value errors translate into end-to-end accuracy loss using three DNN models: a CNN (ResNet-50) and two transformer-based LLMs (Llama-3.1-8B and Llama-3.2-1B), instantiated from Hugging Face~\cite{huggingface_hub} (PyTorch) with BF16 weights and activations. 
For ResNet-50, we report ImageNet-1k top-1 accuracy; for the LLMs, we report ARC-Easy accuracy (four-way multiple choice, so random guessing yields $0.25$).

To simulate the effect of DRAM errors, we \ADD{inject} random bit flips in both the weights and \ADD{intermediate} activations of the networks \ADD{during PyTorch-based inference} (Section~\ref{sec:eval:reliability}).
For each target \emph{Bit Error Rate (BER)}, we inject errors uniformly across all bit positions of BF16 and assess the model’s accuracy.
For each model and BER, we run 100 Monte-Carlo trials and summarize the resulting accuracies using box-and-whisker plots.

\subsubsection{Overall Robustness}

Figure~\ref{fig:dnn_comparison}(a-c) compare the overall robustness of the three models. 
ResNet is relatively tolerant to errors: its accuracy is essentially unchanged up to $\text{BER}=10^{-9}$.
At $\text{BER}=10^{-8}$, the distribution broadens---most runs remain reasonably accurate while some collapse toward zero---indicating that only certain errors are catastrophic. 
By $\text{BER}=10^{-7}$ most runs produce near-zero accuracy, and higher BERs completely destroy the model.

The two LLMs exhibit much higher vulnerability, losing accuracy at BERs roughly two to three orders of magnitude lower than the CNN. 
Llama-3.1 already shows noticeable degradation at $\text{BER}=10^{-10}$, and all trials converge to chance-level performance by $\text{BER}=10^{-9}$. 
Llama-3.2 is slightly more robust but follows the same trend: a non-trivial fraction of runs collapses at $\text{BER}=10^{-10}$ and most do so by $\text{BER}=10^{-9}$.

These results highlight two main observations:

\begin{itemize}
\item{ \textbf{LLMs can be significantly more fragile than CNNs}, likely due to token-based processing and transformer dynamics (attention amplification and residual pathways that can propagate outliers).}
\item{ \textbf{Accuracy varies widely across trials at a fixed BER}, suggesting that a small number of extreme value outliers trigger catastrophic degradation.}
\end{itemize}

\subsubsection{Bit-level Robustness}

We now focus on \emph{bit-level} robustness to validate the analysis presented in Section~\ref{sec:motivation:value_errors}. 
We limit error injection to a single bit position at a time, keeping all other bits error-free. Although we present results for Llama-3.2, similar trends are observed for other models as well.

Figure~\ref{fig:dnn_bitposition} illustrates these findings. 
Errors in sign and mantissa bits do not significantly affect accuracy until $\text{BER}=10^{-5}$, while errors in exponent bits cause catastrophic failures even at much lower BERs, such as $\text{BER}=10^{-12}$. This aligns with the analysis in Section \ref{sec:motivation:value_errors}, where mantissa flips induce $\times{(<1)}$ perturbations, sign flips cause a bounded $\times(-1)$ change, and exponent flips lead to unbounded range changes.

Additionally, a few unlucky bit flips in critical exponent bits are sufficient to severely degrade LLM accuracy. Llama-3.2-1B, with approximately 1B parameters and 156B activations, experiences $\sim$1.5 flipped bits per trial at a $\text{BER}=10^{-11}$ on $e[7]$. Despite this small error rate, 10 out of 100 trials drop to near random-guess accuracy. Even at $\text{BER}=10^{-12}$, which corresponds to an average of 0.15 flipped bits per trial, 1 out of 100 trials still collapses, showing that a \emph{single} unlucky exponent flip can be enough to break the model.


These bit-level results reinforce three key insights:
\begin{itemize}
    \item \textbf{Not all bits are equally important:} Sign and mantissa bits have relatively minor effects, whereas exponent bits—especially high-order ones—dominate failures.
    
    \item \textbf{Few critical flips are enough to break the model:} A \emph{single} unlucky exponent flip can be enough to destabilize the model's performance, even with a low BER.

    \item \textbf{At increased BERs, protection must cover more than just the top exponent bit:} While $e[7]$ is the most vulnerable, other exponent bits also lead to severe accuracy degradation as the BER rises (e.g., $10^{-8}$ in the Llama-3.2 case).
\end{itemize}

Collectively, these observations show that DNNs, especially LLMs, are highly fragile to memory errors and require robust protection to maintain reliable performance.

%% file: sections/4.prior.tex
\section{Prior Work}
\label{sec:prior}

Protecting DNN accuracy under memory errors has followed two broad paths. Industry predominantly deploys bit-precise ECC within a fixed redundancy budget\ADD{, whereas academia increasingly explores semantics-aware protection}, leveraging value distributions and model tolerance to extract more robustness per parity bit.

\subsection{Industry Practice}

Modern GPUs provision 16 bits of ECC per 256-bit block~\cite{standard2023high, standard2025high}.
This budget is typically allocated to either bit-level SEC–DED or \emph{Single-Symbol Correction (SSC)} using 8-bit symbol RS codes.
SEC–DED guarantees correction of any single-bit error and detection of any double-bit error within the 272-bit codeword (256 data + 16 parity). Conceptually, it designates \emph{correction} for exactly the 272 single-bit error syndromes out of the $2^{16}=65{,}536$ possible syndromes, and flags most other nonzero syndromes as detected uncorrectable errors (DUEs). This allocation yields very strong \emph{detection} coverage (about $99.6\%$ for $>2$-bit errors) and correspondingly low \emph{silent data corruption (SDC)} risk.

Alternatively, vendors can encode 8-bit symbols with RS$(34,32)$ to correct a \emph{single} 8-bit symbol error within the block. Because more syndromes are devoted to correction events, residual \emph{detection} coverage for uncorrectable multi-symbol faults is weaker (about $86.8\%$).
In practice, vendors often prefer SEC--DED for its superior detection properties\ADD{~\cite{nvidia2020a100, sullivan2021characterizing, nvidia2022hopper}} and rely on higher-level mechanisms (e.g., checkpoint\slash rollback) to
recover from DUEs. However, recent large-scale LLM training
highlights that such recoveries incur substantial overhead, frequent job interruptions and degraded cluster availability\ADD{~\cite{llama2024hbm3}}.

\subsection{Academic ECC Solutions}

Recent research has proposed various techniques to mitigate the impact of memory errors on AI models.
Unlike conventional ECC schemes, these methods often leverage the statistical redundancy and value distribution of DNN parameters to improve fault tolerance at low cost.
Representative approaches include the following.

Variable Protection (VP)~\cite{catalan2024bitlevelredundancy} exploits the inherent redundancy of neural network weights to enhance resilience against memory faults.
By utilizing consistent bit patterns observed across weight groups, it enables lightweight error recovery with little impact on model accuracy.
Although VP improves bit-level robustness, its protection coverage is limited by the characteristics of the model architecture and data representation.

Qin et al.~\cite{qin2017weightnulling}  study the robustness of DNNs against storage-media errors and propose \emph{Weight Nulling}, which adds a single parity bit per weight by repurposing the least significant bit.
On a parity mismatch, the scheme replaces the corrupted weight with zero rather than correction.
However, its parity check only detects odd-number bit errors and may lose important information when applied to weights with large magnitudes.

Value-Aware Parity Insertion (VAPI)~\cite{lee2022valueaware} targets 8-bit quantized CNN weights.
It observes that most weights lie near zero and adopts sign-magnitude representation so that some higher-order bits (e.g., $b_6$, $b_5$) are rarely used. 
VAPI overwrites those “less important” bits with parity while keeping the overall storage overhead small. 
Using a DEC(64,50) code, it corrects up to two bit errors per 64-bit block of weights without retraining, but it still focuses on stored weights only and assumes 8-bit quantization with specific value distributions.

PoP-ECC~\cite{park2025popecc} introduces a two-tier design that protects virtual parities (VPs) rather than raw weights, and stores only parities-of-parities (PPs) generated from the VPs. Combined with channel-wise quantization, this design provides strong tolerance to multi-bit upsets. However, its focus on exact value recovery constrains how much efficiency it can ultimately achieve.

Despite their differences, these approaches share a common goal: protecting individual data bits. Some exploit intrinsic redundancy in neural parameters or alter data representations to achieve lightweight fault tolerance, often trading universality or coverage for storage efficiency.

\ADD{\subsection{DNN Reliability}}

\ADD{Another line of research studies end-to-end reliability during DNN execution. Hong et al.~\cite{hong2019terminal} showed that hardware-induced errors can still severely affect DNN inference. To evaluate such vulnerabilities systematically, Mahmoud et al.~\cite{mahmoud2022goldeneye} proposed a framework that quantifies the impact of hardware faults across multiple numerical formats. Most of these studies, however, focus on image-classification models.}

\ADD{Reliability analysis for LLMs has only recently emerged. Sun et al.~\cite{sun2025demystifying} characterized the end-to-end impact of memory and compute errors during LLM inference and showed that error propagation in transformer architectures differs from that in image-classification models. Their study identifies vulnerable bit positions, but it does not provide a detailed characterization of the numerical mechanisms that drive this sensitivity.}

\ADD{Several mitigation techniques aim to improve the fault resilience of DNNs. Sullivan et al.~\cite{sullivan2018swapcodes} proposed \emph{SwapCodes}, a hardware--software cooperative mechanism that leverages register-file ECC and instruction duplication to detect transient errors in GPU pipelines. Wasim et al.~\cite{wasim2023hardware} showed that initializing the classification layer with text-guided embeddings can improve the resilience of image classifiers by increasing classification margins and reducing sensitivity to activation faults. Chen et al.~\cite{chen2021ranger} proposed \emph{Ranger}, which constrains layer activations to statistically profiled ranges and suppresses catastrophic fault-induced value spikes.}

%% file: sections/5.main.tex
\section{\mytitle{}}
\label{sec:main}

This section presents \mytitle{}, a novel ECC framework that preserves DNN accuracy under severe memory errors. With the standard HBM ECC budget of 16 parity bits per 256-bit block (6.25\% redundancy), \mytitle{} tolerates more than 64 flipped data bits per block—roughly $8\times$ the raw-bit coverage of conventional bit-centric schemes, which can correct up to 8 flipped bits under the same budget.

Conventional ECC protects individual \emph{data bits}, so redundancy grows with the number of bits being guarded. Some prior designs attempt to save redundancy by protecting only “important” bits (e.g., integer MSBs). For floating-point DNNs, however, this approach is fragile: as shown in Section~\ref{sec:motivation:value_errors}, flips in sign or exponent 
can amplify error magnitude beyond the original value, 
and even the most-significant mantissa bit shifts a value by one-half. Simply prioritizing a subset of bits does not align with what actually drives model failure.

\mytitle{} instead protects compact, per-value \emph{metadata} in the form of \emph{Range Identifiers (RIDs)}. The numeric domain is partitioned into a small number of ranges, and each value is mapped to the index of its range (e.g., a 4-bit RID). Because RIDs are much smaller than the underlying values, they can be strongly protected within a tight parity budget. On a fault, \mytitle{} first corrects the RID and then \emph{approximately} reconstructs the value by substituting a representative of that range, bounding the post-repair error by the range width.

\subsection{Architecture}
\begin{figure}[t]
    \centering
    \includegraphics[width=\columnwidth]{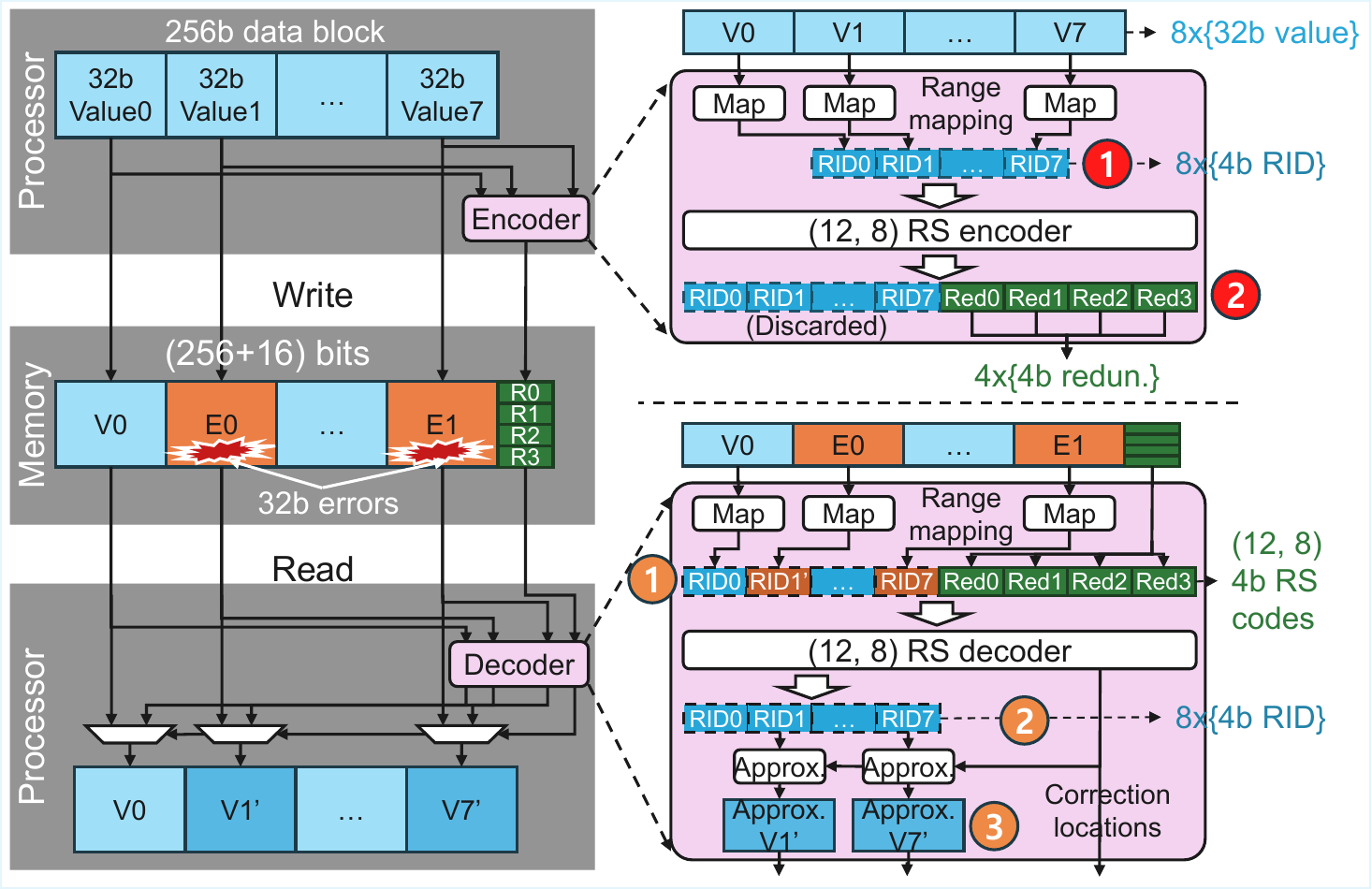}  
    \caption{\mytitle{} on a 256-bit block containing eight FP32 values.}

    \label{fig:main:overview}
    \vspace{-2pt}
\end{figure}

Figure~\ref{fig:main:overview} shows \mytitle{} applied to a 256-bit block with eight FP32 values. In this example, two FP32 values are completely corrupted (Value1$\to$Error0 and Value7$\to$Error1, 64 bits in total).

\subsubsection{Write path (encode)}

Each FP32 value is first mapped to one of 16 predefined ranges (\circnumred{1}), producing a 4-bit RID. The eight RIDs are then encoded using a RS code over 4-bit symbols, RS$(12,8)$, which generates four 4-bit parity symbols (16 bits total) (\circnumred{2}). The encoder discards the explicit RIDs and stores only the original 256-bit data plus the 16-bit parity.

\subsubsection{Memory errors}
During storage or transfer, memory faults can perturb the data. We categorize them as:
\begin{itemize}
  \item \textbf{Intra-range errors:} Bit flips change the value but keep it within the same range (e.g., lower mantissa flips). These are considered semantically benign.
  \item \textbf{Inter-range errors:} Bit flips move the value into a different range (e.g., sign/exponent flips or large mantissa shifts) and are considered harmful.
\end{itemize}

\subsubsection{Read path (decode)}
On a read, \mytitle{} regenerates the eight RIDs from the received FP32 values (\circnumorange{1}); corrupted data may yield erroneous RIDs. The regenerated RIDs and the stored RS$(12,8)$ parity are passed to the RS decoder, which corrects up to two erroneous RID symbols (\circnumorange{2}). For each corrected RID, the corresponding FP32 value is replaced with the representative of that range (\circnumorange{3}), achieving \emph{bounded approximate correction}.

If the RS code can correct $t$ symbols, \mytitle{} tolerates up to $t$ \emph{inter-range} errors (RID errors) per block, while any number of benign \emph{intra-range} errors pass through without consuming correction capacity. 
Each value position uses a small mux: if the RID decoder flags no error \ADD{(either a true no-error case or an intra-range error)}, the raw FP32 value is forwarded; if an inter-range error is corrected, the range representative is used instead. When the number of inter-range errors exceeds $t$, the decoder raises a \emph{DUE} \ADD{(or results in SDC in rare cases)}.

Conceptually, RIDs resemble quantization levels, but with a crucial difference: \mytitle{} uses ranges only as \emph{recovery targets}, not as the primary storage format. Error-free values are stored and used at full precision; only corrupted values are snapped back to their range representatives. This metadata-centric, range-aware design is what allows \mytitle{} to provide strong, bounded-error protection under a fixed GPU ECC budget.

\subsection{Range Mapping}
\label{sec:main:rangemapping}

Range mapping decides (i) how values are grouped into ranges and (ii) how large the post-repair error can be, so it has a direct impact on DNN accuracy. If ranges are too \emph{wide}, \mytitle{} wastes its limited RID space on rarely used magnitudes (e.g., $2^{127}$) and cannot describe common values precisely. If ranges are too \emph{narrow}, many RIDs are spent on almost identical values (e.g., $[2^{-126},2^{-125})$ vs.\ $[2^{-125},2^{-124})$) for only marginal MAE improvement. The goal is to construct a \emph{RangeMap} that balances coverage of realistic value distributions against a tight RID budget and bounded-error requirements.

\subsubsection{Ideal Mapping}

We define the RangeMap objective as minimizing \emph{Mean Absolute Error (MAE)} after repair:
\begin{equation}
    \mathrm{MAE}
    = \int_{-\infty}^{\infty} PDF(x)\,|x-\hat{x}(x)|\,dx,
\end{equation}
where $PDF(\cdot)$ is the distribution of $x$, and $\hat{x}(x)$ is the representative of the range that $x$ falls into.

If $x$ follows a normal distribution with mean $0$ and standard deviation $\sigma$, i.e., $x \sim \mathcal{N}(0,\sigma^2)$, and we use 2-bit RIDs (4 ranges) for BF16, we parameterize the four ranges as
\[
(-\infty, c_1\sigma),\; [c_1\sigma, c_2\sigma),\; [c_2\sigma, c_3\sigma),\; [c_3\sigma, +\infty),
\]
with corresponding representatives $r_1\sigma, r_2\sigma, r_3\sigma, r_4\sigma$.

For a given choice of $\{c_i,r_i\}$, the MAE is
\begin{equation}
\label{eq:mae-4range}
    \mathrm{MAE}
    = \sum_{i=1}^{4}
      \int_{c_{i-1}\sigma}^{c_i\sigma}
        PDF(x)\,|x - r_i\sigma|\,dx,
\end{equation}
where $c_0=-\infty$ and $c_4=+\infty$.

We solve this optimization once offline using standard L1-optimal scalar quantization. For the normalized Gaussian, the optimal thresholds and representatives are
\[
(c_1,c_2,c_3)=(-0.8217,\,0,\,0.8217)
\]
\[
(r_1,r_2,r_3,r_4)=(-1.2657,\,-0.3778,\,0.3778,\,1.2657).
\]
These normalized values can be reused for any format (e.g., FP32, BF16, INT8) whose data follow a similar distribution after scaling by $\sigma$.

\subsubsection{Simple Mapping}
\begin{figure}[t]
    \centering
    \includegraphics[width=0.95\columnwidth]{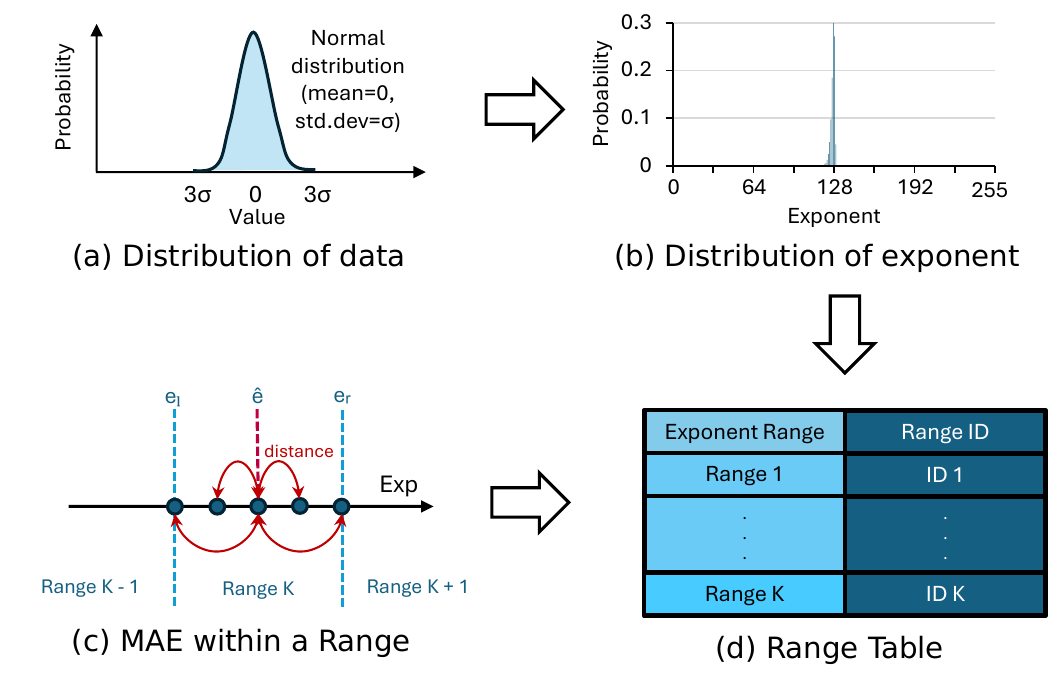}
    \caption{Construction algorithm for simple range mapping.}
    \label{fig:main:range_mapping}
    \vspace{-2pt}
\end{figure}

In practice, \mytitle{} may apply several RangeMaps in parallel within a block and support multiple formats at once. To keep hardware simple, we restrict the mapping logic to the exponent field (motivated by Section~\ref{sec:motivation:value_errors}). For BF16, the construction procedure is summarized in Figure~\ref{fig:main:range_mapping}.

We first derive the exponent distribution for the same distribution of a zero-mean Gaussian with standard deviation $\sigma$.
Let $e$ denote the 8-bit BF16 exponent. Its probability mass function is
\begin{equation}
P(e=k)
= 2\!\left[
\Phi\!\left(\frac{2^{\,k+1-127}}{\sigma}\right)
-\,
\Phi\!\left(\frac{2^{\,k-127}}{\sigma}\right)
\right],
\end{equation}
where $\Phi(\cdot)$ is the Cumulative Distribution Function (CDF) of the standard normal distribution.

Next, we build a range table over the exponent domain that minimizes MAE at the \emph{magnitude} level. Let the exponent space be
\[
\mathcal{E} = \{0, 1, \dots, 255\},
\]
and define a monotonic mapping function representing the numerical scale of each exponent:
\[
f(e) = 2^{e-127}.
\]

We partition $\mathcal{E}$ into $K$ contiguous intervals
$\{[l_k, r_k]\}_{k=1}^{K}$,
and assign a representative exponent $\hat{e}_k \in [l_k, r_k]$ to each interval.
In this case, all values within a given interval are bounded by the same representative exponent after \mytitle{} recovery.
Therefore, the overall MAE can be expressed as the weighted sum of reconstruction errors within each range:
\[
L = \sum_{k=1}^{K}\sum_{e=l_k}^{r_k} P(e)\,|\,f(e) - f(\hat{e}_k)\,|.
\]
Through this process, we obtain a \emph{globally optimal range table} 
that minimizes the overall MAE under the given exponent distribution 
and average error model. 

Table~\ref{tab:range_example} shows an example 4-entry range table for $\sigma = 4$. This $\sigma$ is conservatively chosen so that most LLM values (roughly $\pm 3\sigma \approx \pm 12$) fall within the covered region.

\begin{table}[t!]
\centering
\caption{Example of a 4-entry range table when $\sigma$ = 4.}
\footnotesize
\renewcommand{\arraystretch}{1.1}
\begin{tabular}{c c c c}
\Xhline{2\arrayrulewidth}
\textbf{\stackanchor{Exponent}{Range}} &
\textbf{\stackanchor{Value}{Range}} &
\textbf{\stackanchor{Boundary}{Exponent}} & \textbf{\stackanchor{Representative}{Value}}\\
\hline
[0, 127]   & (0, 2) & 01111111 & 0.5 \\ 
128        & [2, 4) & 10000000 & 2 \\
129        & [4, 8) & 10000001 & 4 \\
{}[130, 255] & [8, $\infty$) & 11111111 & 8  \\
\Xhline{2\arrayrulewidth}
\end{tabular}
\label{tab:range_example}
\end{table}

\subsection{ECC Configurations}
\label{sec:main:eccconfig}

\mytitle{} is a flexible framework that can be tuned to different reliability targets under a fixed 16-bit parity budget.
We consider two RS–based designs that trade off correction strength and reconstruction fidelity:
\emph{8b Single-Symbol Correction (SSC)} and \emph{4b Double-Symbol Correction (DSC)}, shown in Figure~\ref{fig:main:encoding}.

In both cases, protection is organized at 32-bit granularity: each 32-bit value corresponds to one RS symbol, so common burst faults (e.g., 16-bit local wordline faults or 32-bit sub-wordline driver faults) fall within a single symbol and can be corrected with less redundancy.

\subsubsection{8b SSC}
The 8b SSC configuration uses 8-bit symbols and an RS(10,8) code, where 8 data symbols and 2 parity symbols form one ECC word.
A 256-bit block is split into eight 32-bit values, each mapped to an 8-bit RID symbol; the 16-bit ECC lane stores two parity symbols and enables correction of any single symbol in the block (up to 32 corrupted data bits).
This configuration provides fine-grain range information and thus higher-quality approximate reconstruction.

\subsubsection{4b DSC}
The 4b DSC configuration uses 4-bit symbols with an RS(12,8) code, comprising 8 data symbols and 4 parity symbols.
Each 32-bit value is mapped to a 4-bit RID symbol; the 16-bit ECC lane holds four parity symbols and supports correction of up to two symbols (up to 64 corrupted data bits).
Because each value gets fewer RID bits, the recovered values are coarser, but the configuration offers stronger protection against overlapping faults.
The 12-symbol word length respects the RS maximum of 15 symbols for 4-bit symbols.
\begin{figure}[t]
    \centering
    
    \subfloat[8-bit RS(10, 8) Single Symbol Correction for moderate BERs.]{
        \includegraphics[width=.9\columnwidth]{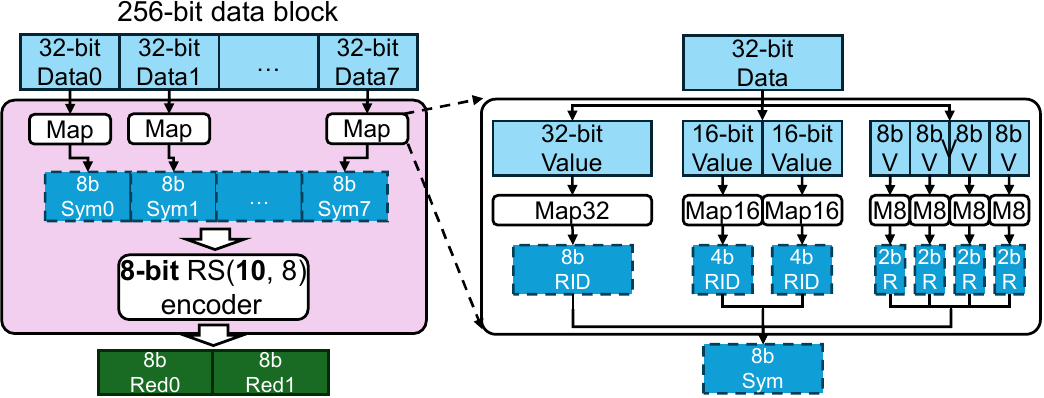}
        \label{fig:main:encoding1}
    }
    
    \subfloat[4-bit RS(12, 8) Double Symbol Correction for high BERs.]{
        \includegraphics[width=.9\columnwidth]{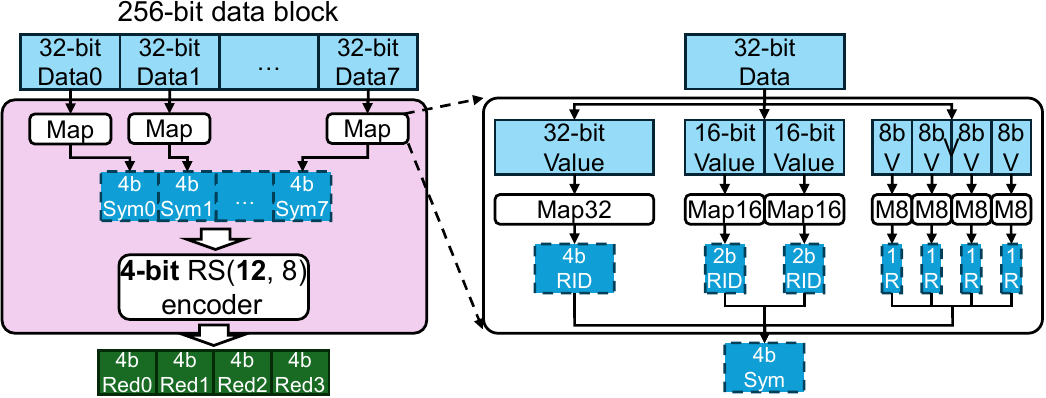}
        \label{fig:main:encoding2}
    }
    \caption{\mytitle{} ECC configurations.}
    \label{fig:main:encoding}
    \vspace{-2pt}
\end{figure}

\subsection{\ADD{Multi-Format Support}}

Modern DNNs employ multiple numeric formats (e.g., FP32, BF16, FP8, INT8) and exhibit distinct value distributions across layers (weights vs.\ activations, early vs.\ late blocks). \mytitle{} adapts to this diversity in a structured way.
First, \mytitle{} can host multiple RangeMaps in parallel, each tuned to a specific format and tensor class (e.g., FP32 weights, BF16 activations, FP8 logits). As shown in Figures~\ref{fig:main:encoding1} and \ref{fig:main:encoding2}, each RangeMap instance includes sub-maps for 32-, 16-, and 8-bit values. 
For sub-32b values, each sub-map produces smaller RIDs, and several of these are packed into a single \emph{RID symbol} to match the ECC symbol width.
As the value size shrinks, more values fit in 32-bit data, so we increase the number of sub-maps and reduce RID bits per value.
This allows the same ECC budget to be reused efficiently across formats without assigning ranges to values that never appear. On the downside, this incurs more hardware costs by requiring multiple instances of sub-maps per 32-bit data. 

\ADD{We choose the per-value RID widths to satisfy both semantic fidelity and DRAM fault characteristics. In particular, common DRAM fault modes can corrupt up to 32 aligned bits within a block (Section~\ref{sec:motivation:errors}). To make such a fault appear as a single symbol error, \mytitle{} packs the RIDs of values that share the same 32-bit-aligned region into one RID symbol. A 32-bit region can contain two 16-bit values, four 8-bit values, or eight 4-bit values. For example, using 4/2/1-bit RIDs for 16/8/4-bit values lets all RIDs in a 32-bit region fit into one 8-bit symbol, enabling correction with a single 8-bit-symbol SSC. Alternatively, using 2/1-bit RIDs for 16/8-bit values packs the region into one 4-bit symbol, enabling correction with a single 4-bit-symbol code while leaving additional correction capability for multiple faults.}

Second, the memory controller selects the appropriate sub-map using a small tag embedded in the physical address\ADD{, similar to prior address-based ECC selection schemes~\cite{yoon2010virtualized}. Specifically,} \mytitle{} repurposes unused upper address bits \ADD{(e.g., bit[57:54])} as a \emph{Map Tag}, enabling the controller to identify the target sub-map without extra pins or sideband signals and to switch mappings on the fly. On reads, the decoder consults this tag to pick the correct sub-map and reconstruct the representative value from the decoded RID.
\ADD{For data regions that require exact recovery, the Map Tag can also select a conventional ECC scheme (e.g., SEC-DED), thereby providing exact but weaker protection. This mechanism allows \mytitle{} to support different data types and protection requirements with minimal hardware changes.}

Taken together, these components---RID-based architecture, SSC/DSC ECC configurations, exponent-aware RangeMaps, and multi-format support—turn \mytitle{} into a drop-in, GPU-compatible protection layer that spends the ECC budget on what actually matters: keeping values in safe numeric ranges. The memory system still stores and moves raw data at full precision, but when faults occur, \mytitle{} detects harmful range changes, snaps corrupted values back to well-chosen representatives, and lets benign noise pass through. In the next section, we show that this design yields strong coverage under realistic DRAM fault modes, preserves CNN and LLM accuracy at high BERs, and incurs negligible area and performance overhead.

%% file: sections/6.evaluation.tex
\newcolumntype{L}[1]{>{\raggedright\arraybackslash}m{#1}}
\newcolumntype{R}[1]{>{\raggedleft\arraybackslash}p{#1}}
\newcolumntype{C}[1]{>{\centering\arraybackslash}m{#1}}

\section{Evaluation}
\label{sec:eval}

This section evaluates the reliability benefits and hardware costs of \mytitle{} compared to existing state-of-the-art error correction schemes.

\subsection{Error Coverage}
\label{sec:eval:coverage}

\begin{table}[t!]
\centering
\caption{A comparison of correction and detection coverage in various fault scenarios.}
\footnotesize
\renewcommand{\arraystretch}{0.95}

\begin{threeparttable}
\resizebox{\columnwidth}{!}{
\begin{tabular}{ l | c | r | r | r | r | r }

\Xhline{3\arrayrulewidth}

\makecell{Fault \\ Modes} &
\makecell{Result} &
\ {Baseline} & 
\makecell{Weight\\Nulling} &
\makecell{VAPI} &
\makecell{RG 8b \\ SSC} &
\makecell{RG 4b \\ DSC} \\

\Xhline{2\arrayrulewidth}

\multirow{2}{*}{SE}
& CE  & \ 100.000 & 0.000   & 100.000 & \multirow{11}{*}{\shortstack[c]{BE\\100.000}} & \multirow{27}{*}{\shortstack[c]{BE\\100.000}} \\
& DUE & \ 0.000   & 100.000 & 0.000   & & \\
\cline{1-5}

\multirow{3}{*}{DAE}
& CE  & \ 100.000 & 0.000   & 100.000 & & \\
& DUE & \ 0.000   & 74.997  & 0.000   & & \\
& SDC & \ 0.000   & 25.003  & 0.000   & & \\
\cline{1-5}

\multirow{3}{*}{16E}
& CE  & \ 100.000 & 0.000   & 0.209   & & \\
& DUE & \ 0.000   & 50.011  & 99.782  & & \\
& SDC & \ 0.000   & 49.989  & $9\!\times\!10^{-3}$ & & \\
\cline{1-5}

\multirow{3}{*}{32E}
& CE  & \ $3\!\times\!10^{-3}$ & 0.000   & $9\!\times\!10^{-6}$ & & \\
& DUE & \ 99.996  & 25.007  & 99.994  & & \\
& SDC & \ $1\!\times\!10^{-3}$ & 74.993  & $6\!\times\!10^{-3}$ & & \\
\cline{1-6} 

\multirow{4}{*}{\makecell[l]{SE+\\SE}}
& CE  & \ 4.949  & 0.000   & 100.000 &          & \\
& BE  &                          &         &         & 94.507   & \\
& DUE & \ 95.051 & 94.117  & 0.000   & 5.493    & \\
& SDC & \ 0.000  & 5.883   & 0.000   & 0.000    & \\
\cline{1-6}

\multirow{4}{*}{\makecell[l]{SE+\\DAE}}
& CE  & \ 4.633  & 0.000   & 93.899  &          & \\
& BE  &                          &         &         & 94.396   & \\
& DUE & \ 95.367 & 72.245  & 6.101   & 5.604    & \\
& SDC & \ 0.000  & 27.755  & 0.000   & 0.000    & \\
\cline{1-6}

\multirow{4}{*}{\makecell[l]{SE+\\16E}}
& CE  & \ 0.000  & 0.000   & 0.173   &          & \\
& BE  &                          &         &         & 88.379   & \\
& DUE & \ 99.999 & 50.001  & 99.826  & 11.621   & \\
& SDC & \ $1\!\times\!10^{-3}$ & 49.999  & $1\!\times\!10^{-3}$ & 0.000 & \\
\cline{1-6}

\multirow{4}{*}{\makecell[l]{SE+\\32E}}
& CE  & \ 0.000  & 0.000   & $1\!\times\!10^{-5}$ &      & \\
& BE  &                          &         &                      & 75.097 & \\
& DUE & \ 99.999 & 25.003  & 99.999  & 24.903   & \\
& SDC & \ $2\!\times\!10^{-3}$ & 74.997  & $8\!\times\!10^{-4}$ & 0.000  & \\
\cline{1-7}

\multirow{4}{*}{FC}
& CE  & \ 0.000  & 0.000   & 0.000   &  & \\
& BE  &                          &         &         & 0.000 & $1\!\times\!10^{-4}$ \\
& DUE & \ 99.998 & $2\!\times\!10^{-3}$  & 100.000 & 99.998 & 99.998 \\
& SDC & \ $2\!\times\!10^{-3}$ & 99.998  &  0.000  & $2\!\times\!10^{-3}$ & $2\!\times\!10^{-3}$ \\
\Xhline{3\arrayrulewidth}
\end{tabular}
}
\end{threeparttable}
\label{table:Coverage}
\end{table}

We begin by comparing the correction capabilities of \mytitle{} against common fault modes encountered in DRAM systems. These fault modes were derived from prior DRAM error studies~\cite{ryu202316,2025dramfaultclassification} and include:
\begin{itemize}
    \item \textbf{Single-Error (SE) Fault:} A fault that causes a single bit to flip within a block, typically resulting from faulty cells or marginal bitlines.
    \item \textbf{Double Adjacent Error (DAE) Fault:} A fault that induces two adjacent bit errors within a block, often occurring due to cell-to-cell bridging faults or TSV faults.
    \item \textbf{16-bit Error (16E) Fault:} A fault that causes up to 16 bit errors within a 16-bit boundary, which can result from issues like local wordline or column select line faults.
    \item \textbf{32-bit Error (32E) Fault:} A fault that leads to up to 32 bit errors within a 32-bit boundary, typically caused by sub-wordline driver faults.
    \item \textbf{Full-Chip (FC) Fault:} A fault that corrupts all bits within a block.
\end{itemize}

\ADD{Because \mytitle{} can tolerate a single instance of most fault modes except FC, we evaluate both single-fault and double-fault cases within a block. For each fault scenario, we randomly select the fault location(s) within a block and inject bit corruptions within the affected region. Except for SE, each bit within the fault boundary is flipped independently with 50\% probability. We then apply the ECC decoder and classify each outcome as a \emph{Corrected Error (CE)}, DUE, or SDC. For \mytitle{}, we report \emph{Bounded Error (BE)} instead of CE because the design targets bounded approximate recovery rather than exact bitwise correction. We repeat each experiment $10^9$ times and report the resulting outcome probabilities.}

\ADD{We compare two configurations of \mytitle{} (8-bit SSC and 4-bit DSC) against a baseline, Weight Nulling~\cite{qin2017weightnulling}, and VAPI~\cite{lee2022valueaware}.
Our baseline models an HBM3-style stack that uses a 16-bit RS(19,17) code as O-ECC to correct storage-side faults inside the device and a 16-bit cyclic redundancy check (CRC16) as S-ECC to provide end-to-end error detection. We assume all errors occur during storage to go through O-ECC and S-ECC.}
For Weight Nulling and VAPI, which repurpose some unused exponent bits for parity, we assume that the value distribution allows for this repurposing.

Table \ref{table:Coverage} compares error coverage across the schemes. 
The baseline provides exact correction only when faults remain within the O-ECC correction range. Its CE coverage is already negligible for 32E, correcting only $3\!\times\!10^{-3}\%$ of cases, and it completely loses CE capability once the fault pattern exceeds the O-ECC range, as in SE+16E and SE+32E.

\begin{table}
\centering
\caption{Fault-mode distribution and expected flipped bits used during the DNN accuracy measurement.}
\footnotesize

\begin{threeparttable}
\begin{tabular}{ L{1.4cm} | R{1.7cm} | R{1.7cm} | R{1.9cm} }

\Xhline{3\arrayrulewidth}

{\centering Fault Mode} &
{\centering Fault Ratio (A)} &
{\centering Error Count (B)} &
{\centering Error Ratio (A$\times$B)} \\

\Xhline{2\arrayrulewidth}

SE  & $0.009\times BER$  & 1  & $0.009\times BER$ \\
DAE & $0.012\times BER$  & 2  & $0.023\times BER$ \\
16E & $0.022\times BER$  & 8  & $0.175\times BER$ \\
32E & $0.050\times BER$  & 16 & $0.793\times BER$ \\

\Xhline{2\arrayrulewidth}
Total & & & $BER$ \\
\Xhline{3\arrayrulewidth}
\end{tabular}

\end{threeparttable}

\label{table:fault_rate}
\end{table}

\begin{figure*}[t]
    \centering
    \subfloat[ResNet-50 / ImageNet-1k top-1 accuracy]{
        \includegraphics[width=0.8\textwidth]{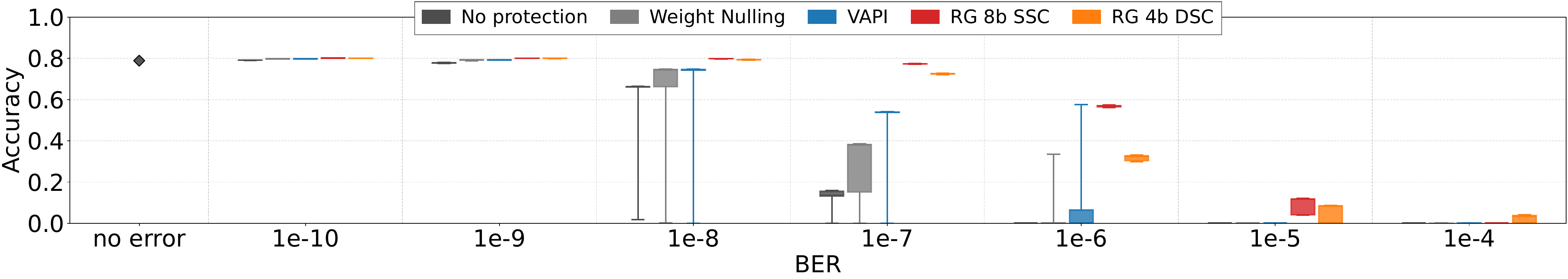}
        \label{fig:resnet_result}
    }
    \vspace{10pt}
    \subfloat[Llama-3.1-8B / ARC-Easy score]{
        \includegraphics[width=0.8\textwidth]{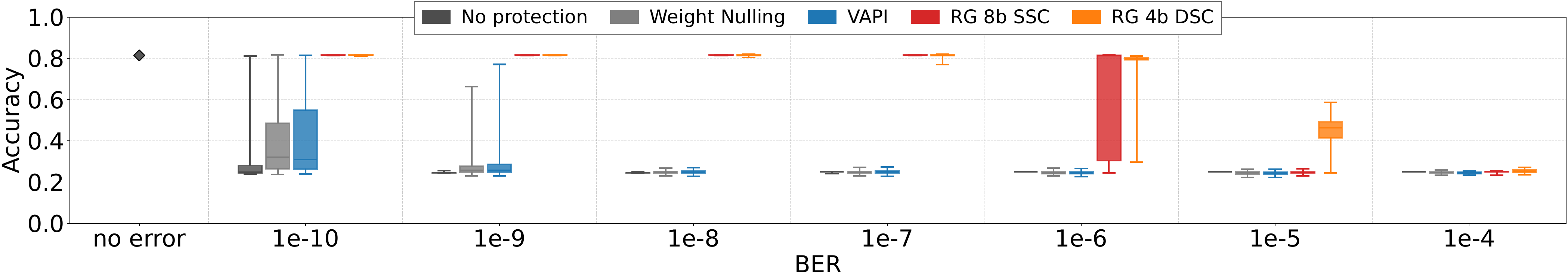}
        \label{fig:llama-8B_result}
    }
    \vspace{10pt}
    \subfloat[Llama-3.2-1B / ARC-Easy score]{
        \includegraphics[width=0.8\textwidth]{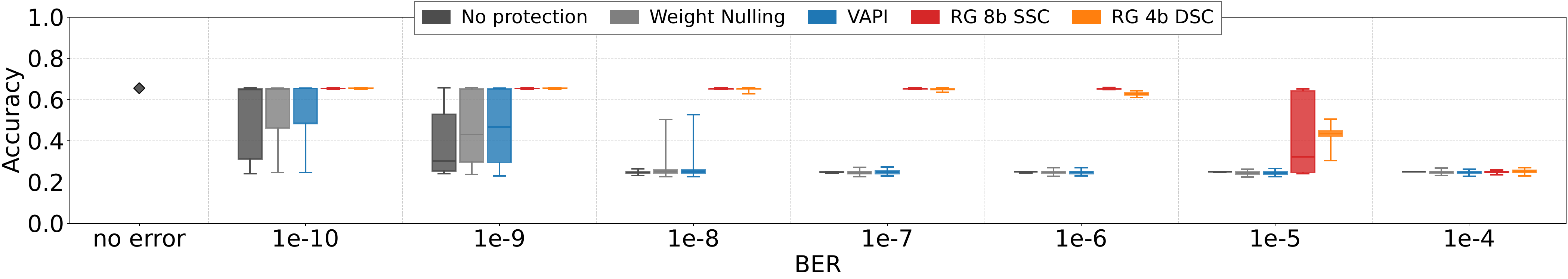}
        \label{fig:llama-1B_result}
    }
    \caption{DNN accuracy degradation with varying BERs.}
    \label{fig:reliability_results}
\end{figure*}

Weight Nulling uses parity bits only for detection and offers no correction; its parity-based checks miss most multi-bit patterns. VAPI corrects SE, SE+SE, and DAE well, but its bit-level protection is still weak against wider multi-bit faults.

In contrast, \mytitle{} provides the strongest overall coverage. RG 8b SSC corrects all single faults (except FC), while RG 4b DSC extends support to two simultaneous faults. Both configurations show negligible misdetection, with an FC misdetection rate of $2\!\times\!10^{-3}\%$. Thus, \mytitle{} is robust for both common (single-fault) and rare (two-fault) scenarios; faults that trigger more than three range changes are expected to be exceedingly rare.

\subsection{DNN Accuracy}
\label{sec:eval:reliability}

We next compare DNN accuracy across schemes under more realistic DRAM fault patterns.
Unlike the single-bit injection experiments in Section~\ref{sec:motivation:dnn_accuracy}, here we randomly inject SE, DAE, 16E, and 32E faults to reflect common post–on-die-ECC failures.
We adopt fault ratios from a DDR5 fault classification study~\cite{2025dramfaultclassification} and weight each mode by its expected number of flipped bits so that the combined contribution of all modes matches the target BER (Table~\ref{table:fault_rate}).
This setup lets us stress each scheme under a \ADD{more} realistic mix of sparse and bursty multi-bit errors rather than idealized single-bit flips.

\ADD{During PyTorch-based inference, we inject randomly generated faults into stored weights and intermediate activations before subsequent layers consume them, following the methodology of recent DNN fault-injection frameworks~\cite{sun2025demystifying, mahmoud2020pytorchfi, wang2024mrfi, realm2025, fidelity2020}. For each fault, we flip bits within the affected region with 50\% probability, apply the corresponding ECC algorithm to the corrupted block, and update the tensor only when the outcome is uncorrectable or bounded-corrected. We then measure the final inference accuracy and repeat the experiment 100 times for each BER and protection scheme. For the LLM experiments, we conduct inference with lm-evaluation-harness~\cite{eval-harness}.}

Figure~\ref{fig:reliability_results} shows how accuracy degrades as BER increases.
The \emph{no-protection} case behaves similarly to the single-bit experiments: ResNet-50, Llama-3.1, and Llama-3.2 exhibit sharp accuracy drops around BERs of $10^{-8}$, $10^{-10}$, and $10^{-10}$, respectively, and quickly approach chance-level performance beyond those points.
For the LLMs, the run-to-run spread is noticeably worse than in the single-bit case, likely because clustered faults (16E, 32E) more frequently strike exponent fields and generate extreme outliers that dominate attention and residual paths.

Weight Nulling and VAPI provide only limited improvement over this baseline.
Their weak coverage for bursty multi-bit errors leaves substantial accuracy loss at moderate BERs, with curves that remain close to the no-protection case, especially for LLMs.

In contrast, \mytitle{} closely tracks the no-error baseline across a wide BER range despite the presence of multi-bit errors.
ResNet and Llama-3.1 maintain near-baseline accuracy up to BER $=10^{-7}$, and Llama-3.2 remains robust up to BER $=10^{-6}$, significantly extending the usable BER envelope compared to other schemes.
Even beyond these thresholds, \mytitle{} degrades much more gracefully than bit-centric approaches, confirming that bounding range changes rather than raw bit mismatches is an effective strategy for preserving DNN and LLM accuracy under realistic fault patterns.
Between the two \mytitle{} configurations, the better choice is model-dependent: 8b SSC outperforms 4b DSC on ResNet and Llama-3.2 by providing more accurate approximate correction with its 16-entry RangeMap, while 4b DSC is preferable for Llama-3.1 thanks to its stronger tolerance to overlapping faults and higher symbol-level correction capability.

\subsection{Hardware Overheads}

\begin{table}[t]
\centering
\caption{\ADD{Hardware overheads of BF16 \mytitle{}}}
\begin{tabular}{@{}l|l|R{2.4cm}|R{2.4cm}@{}}
\Xhline{3\arrayrulewidth}
Metric & Component & RG 8b SSC & RG 4b DSC \\
\Xhline{2\arrayrulewidth}
Area & Encoder (total)      & 3,300 $\mu m^2$ (6,500)   & 800 $\mu m^2$ (1,600) \\
     & \quad - RangeMaps    & 3,100 $\mu m^2$ (6,200)   & 650 $\mu m^2$ (1,300) \\
     & \quad - ECC encoder  & 200 $\mu m^2$ (300)       & 150 $\mu m^2$ (300)     \\
\cline{2-4}
     & Decoder (total)      & 7,800 $\mu m^2$ (15,400)  & 4,200 $\mu m^2$ (8,200) \\
     & \quad - RangeMaps    & 6,700 $\mu m^2$ (13,200)  & 1,600 $\mu m^2$ (3,000) \\
     & \quad - ECC decoder  & 1,100 $\mu m^2$ (2,200)   & 2,600 $\mu m^2$ (5,200)  \\
\Xhline{2\arrayrulewidth}
Power & Encoder (total) & 0.54 mW & 0.18 mW \\
      & Decoder (total) & 0.61  mW & 0.63 mW \\
\Xhline{3\arrayrulewidth}
\end{tabular}
\label{table:hardware_overhead}
\end{table}

\ADD{To estimate the hardware cost of \mytitle{}, we implement its encoder and decoder in SystemVerilog and evaluate two configurations. For RG 8b SSC, the encoder extracts 8-bit RIDs using a flip-flop-based RangeMap. For each 32B access, it compares the 8-bit exponent of each of the sixteen 16-bit values against all 16 RangeMap entries in parallel, requiring 256 8-bit comparators but completing RID extraction in one cycle. A second cycle generates the redundancy bits. RG 4b DSC uses the same structure with a smaller 4-entry RangeMap, reducing the lookup to 64 8-bit comparators, and uses a different generator matrix for redundancy generation.}

\ADD{The decoder regenerates RIDs from the received data in one cycle using the same RangeMap-based lookup structure as the encoder. For area reporting, we model a separate RangeMap in the decoder, although the encoder and decoder can share a single flip-flop-based table in practice. In the next cycle, the decoder computes the syndrome and performs error detection. When an error is detected, it applies correction  using a modified Berlekamp--Massey algorithm hardened in logic gates~\cite{chien2003cyclic,forney2003decoding,liang2017efficient}, requiring one additional cycle per correction.}

\ADD{We synthesize the RTL using Synopsys Design Compiler with a UMC 28nm standard-cell library at 1GHz. We report area in NAND2 equivalents to make the results more process independent, and we further break down the area into RangeMap and ECC logic to help estimate the overhead of configurations with multiple RangeMaps.}

\ADD{Table~\ref{table:hardware_overhead} presents the results. The combined area of the 8-bit SSC encoder and decoder is 11,100~$\mu\mathrm{m}^{2}$, which is larger than that of the 4-bit DSC configuration (5,000~$\mu\mathrm{m}^{2}$) due to its larger RangeMap. Nevertheless, this total overhead corresponds to only \ {21,900} NAND2 equivalents (\ {87,600} transistors). Considering modern GPUs with billions of transistors this is a small portion of the die area (e.g., 5e-7 in 208B-transistor Blackwell).}

\subsection{GPU System Performance}
\label{sec:eval:performance}

\begin{figure}[t!]
  \centering
  \vspace{-10pt}
  \includegraphics[width=1.0\linewidth]{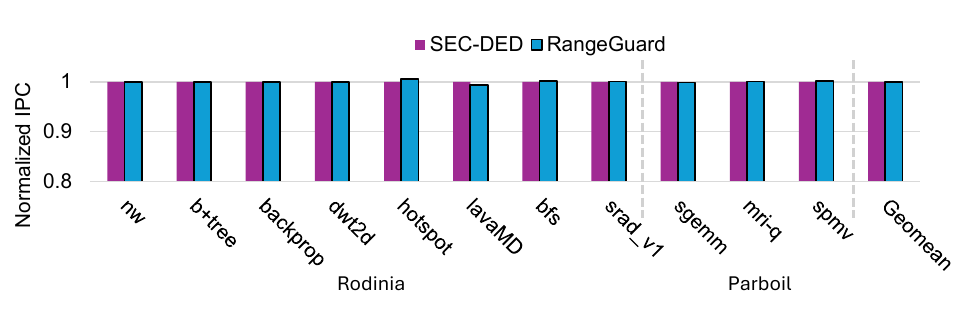}
  \vspace{-10pt}
  \caption{IPC comparison between SEC-DED and \mytitle{}.}
  \label{fig:performance}
  \vspace{-5pt}
\end{figure}

\ADD{To assess the performance cost of the protection, we use the cycle-level GPU simulator Accel-Sim~\cite{khairy2020accel}, configured to model an NVIDIA V100 GPU. 
Since memory errors are rare (occurring on the scale of days or months), we focus on evaluating \mytitle{}'s impact in error-free scenarios.}

\ADD{We adjust the read latency to reflect the decoding overhead. We assume that \mytitle{} requires 2 cycles to return error-free data (immediately after error detection stage), while the baseline already spends 1 cycle on error detection, resulting in a net 1-cycle increase in read latency. We do not increase the write latency, because the baseline already uses 1 cycle for encoding and the additional cycle can be hidden while the request waits in the memory-controller queue or during DRAM write latency (i.e., the interval between the DRAM write command and the arrival of write data). We evaluate 11 workloads from Rodinia~\cite{che2010characterization} and Parboil~\cite{stratton2012parboil}}.



Figure~\ref{fig:performance} shows the results. Owing to the GPU’s throughput-oriented design, the extra cycle of read latency has a negligible effect on performance, reducing geomean \ADD{Instructions-Per-Cycle (IPC)} by only 0.008\%. Even for the most memory-intensive workloads in our suite, the IPC drop remains below 0.05\%, while compute-bound workloads are virtually unaffected. These results indicate that \mytitle{}’s protection can be deployed in GPU systems without sacrificing throughput and system performance.

{
\subsection{Sensitivity Analysis}

\begin{figure}[t]
    \centering
    \subfloat[Sensitivity to $\sigma$ value]{
        \includegraphics[width=0.62\columnwidth]{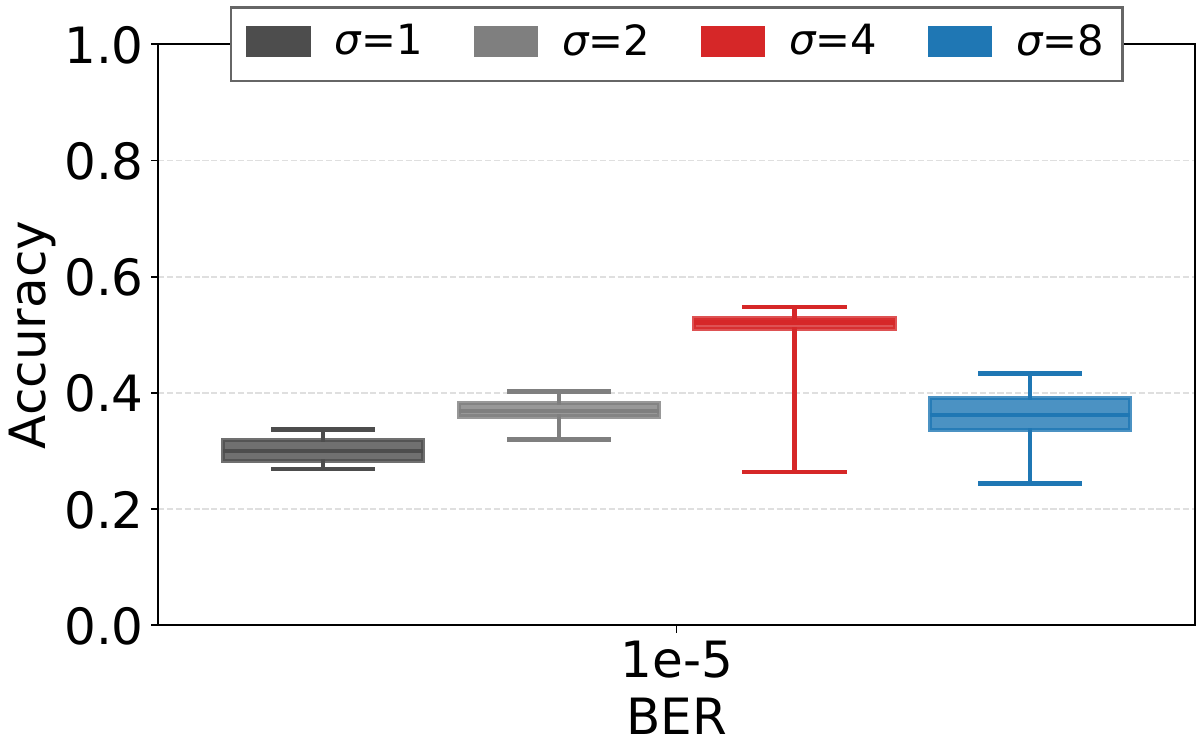}
        \label{fig:eval_sigma_sens}
    }
    \hfill 
    \subfloat[Global vs. Tensorwise]{
        \includegraphics[width=0.31\columnwidth]{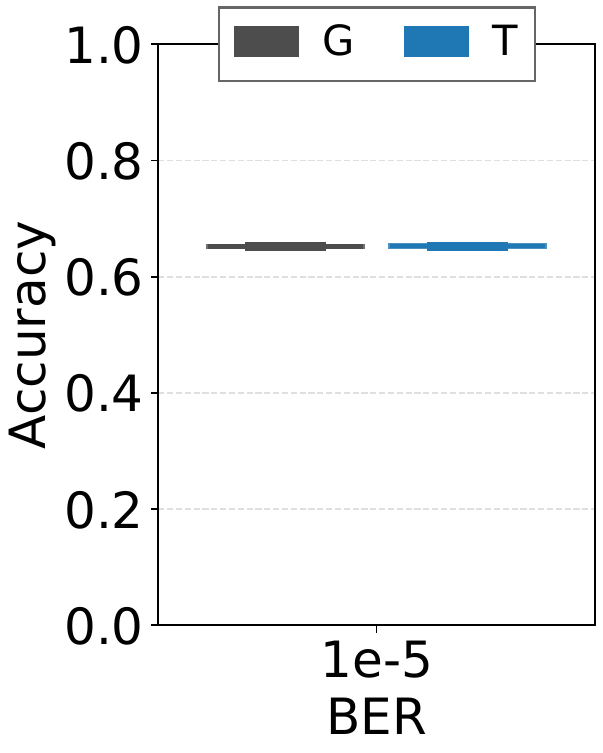}
        \label{fig:eval_fine_grained}
    }

    \vspace{-5pt}
    \caption{Experimental results on robustness and hyperparameter sensitivity.}
    \label{fig:overall_eval_optimized}
    \vspace{-10pt}
\end{figure}

\ADD{To study how semantic protection in \mytitle{} depends on mapping parameters, we conduct a sensitivity analysis of range mapping by varying the 
sigma value and mapping granularity, using the same experimental methodology as in Section~\ref{sec:eval:reliability}.}

\textbf{Sigma Size:} To evaluate the sensitivity of the $\sigma$ value, we first constructed a RangeMap using the simple mapping scheme described in Section~\ref{sec:main:rangemapping} while using 2-bit RIDs. As discussed in Section~\ref{sec:main:rangemapping}, $\sigma$ plays a critical role in balancing parameter coverage against mapping resolution. Fig.~\ref{fig:eval_sigma_sens} highlights that protection efficacy is highly dependent on an optimal $\sigma$. Both excessively small and large $\sigma$ values fail to adequately capture the value distribution, thereby undermining the semantic protection and triggering a sharp decline in model accuracy.

\textbf{Mapping Granularity:} We compared global mapping (G) and tensorwise mapping (T) to evaluate the influence of mapping granularity on diverse internal distributions (Fig.~\ref{fig:eval_fine_grained}). Global mapping uses a fixed configuration ($\sigma = 4$, 4-bit RID), and tensorwise mapping generates dedicated RangeMaps via the Lloyd-Max algorithm~\cite{max1960quantizing, lloyd1982least} modified to minimize MAE. 
While this approach provides more precise reconstruction, it incurs substantial area overhead in the encoder and decoder to manage the massive number of unique RangeMaps.
Despite its simplicity, global mapping provides sufficient resilience. This is because, at a bit error rate of $10^{-5}$, the precision of the global mapping already ensures the semantic integrity of the model without requiring tensor-specific optimization.
}

%% file: sections/7.conclusion.tex
\section{Conclusion}
\label{sec:conclusion}

As DRAM scales and 3D-stacked memories become pervasive, multi-bit faults are increasingly common and especially harmful for modern DNNs and LLMs. We show that a few exponent flips can create extreme outliers, collapsing accuracy at very low BERs where prior bit-centric mitigation schemes already break down.

This paper presented \emph{\mytitle{}}, a metadata-centric framework that protects \emph{ranges} instead of raw bits. \mytitle{} encodes compact Range Identifiers (RIDs) into the existing 16-bit parity bits per 32-byte block, detects \emph{inter-range} deviations, and performs bounded approximate correction by restoring the range and substituting a representative value. We develop MAE-aware RangeMap construction and two concrete configurations (8b SSC, 4b DSC) that fit within current GPU memory interfaces. Under realistic DRAM fault modes, \mytitle{} tolerates errors affecting 64+ data bits per block while keeping errors bounded, preserves near-baseline accuracy for CNNs and LLMs at high BERs, and incurs only a negligible performance impact (0.008\% IPC loss). These results demonstrate that reallocating limited redundancy from bit correctness to semantic range consistency is a practical and effective direction for building fault-resilient DNN/LLM systems on future error-prone memories.